\documentclass[prb,preprintnumbers,footinbib,twocolumn]{revtex4}

\usepackage{graphicx}
\usepackage{dcolumn}
\usepackage{bm}
\usepackage{hyperref}
\usepackage{amssymb}

\begin{document}

\title{Optical conditional spin gates in coupled quantum dots using the electron-hole exchange interaction}

\author{Sophia E. Economou and T. L. Reinecke}

\affiliation{Naval Research Laboratory, Washington, DC 20375, USA}
\date{\today}

\begin{abstract}
We propose a fast optically induced two-qubit \textsc{c-phase} gate between two resident spins in a pair of coupled quantum dots. An excited bound state which extends over the two dots provides an effective electron-electron exchange interaction. The gate is made possible by the electron-hole exchange interaction, which isolates a single transition in the system. When combined with appropriate single qubit rotations, this gate generates an entangled state of the two spins.
\end{abstract}

\maketitle

\section{Introduction}

Spins in quantum dots (QD) are currently under intense investigation due to their potential use as bits of quantum information. All-optically controlled QDs are particularly attractive, as they take advantage of the speed at which optical transitions can be accessed with high degree of control. Qubit operations are typically realized through Raman-like transitions via optical generation of an electron-hole pair. Thus far, many of the requirements of quantum computation have been demonstrated experimentally in QD systems: initialization,\cite{xiaodong} readout,\cite{atature,mikkelsen} and long coherence times,\cite{greilich_science} as well as some recent advances in single-qubit control.\cite{berezovsky,ramsay} Recently conditional control between a spin and an exciton qubit was demonstrated in coupled QDs.\cite{robledo} However, two-qubit gates between spins, which are necessary for spin-based quantum computation, are  yet to be realized in those systems. Such gates are conditional, i.e., they alter the state of the one qubit depending on the state of the other. For such an operation, the ability to switch on an interaction between the two qubits is required. Proposals for optical two-spin gates have been made using a cavity photon mode\cite{imamoglu,clark} or using optical excitation of electron-hole pairs \cite{ORKKY,nazir,Lovett,calarco,emary} as mechanisms for this interaction.

\begin{figure}[htp]
\begin{center}
\includegraphics[width=3.8cm,angle=-90]{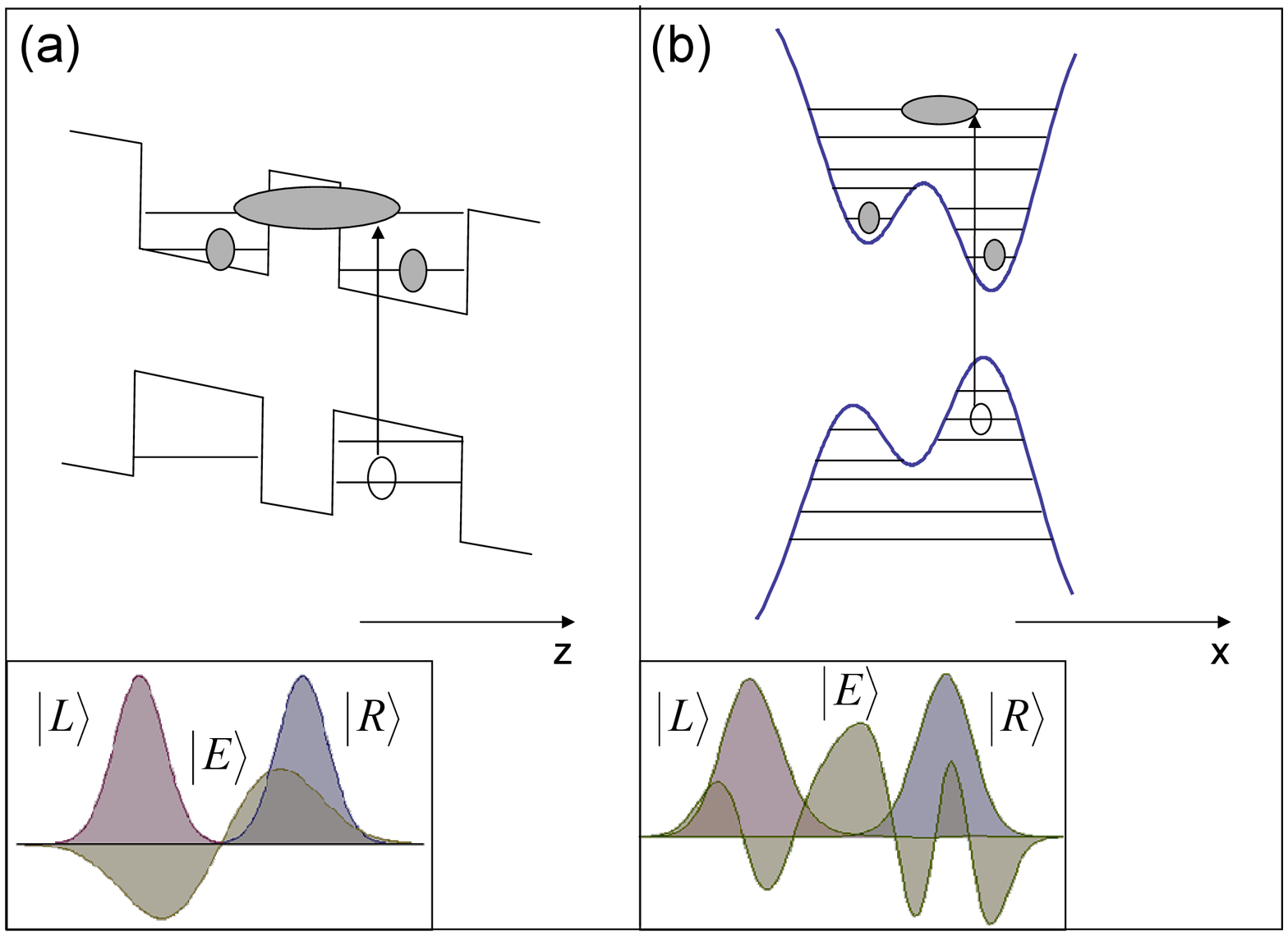}%
\end{center}
\caption{(Color online) Sketch of the potential of (a) laterally and (b) vertically coupled QDs. The single particle orbital states are also sketched, and the arrow denotes the optical transition. The wavefunctions of the three electrons participating in the two-qubit quantum gate are shown in the insets. States $|L\rangle$ and $|R\rangle$ are confined within each QD and have negligible overlap between them. Excited state $|E\rangle$ is the extended bound state which mediates the effective exchange interaction.}
\label{eoverlaps}
\end{figure}

In this work we consider an approach where the intermediate electron is in an excited bound state that extends over both dots. Such states have the advantage of weaker coupling to the continuum, and they have been studied experimentally.\cite{stinaff}  Our scheme is general enough to cover both vertically stacked coupled QDs, and also laterally coupled QDs.\cite{wang_jap,michler} The latter are thought to be desirable for a scalable quantum computing architecture, in which vertically coupled QDs would be used to encode logical qubits in multiple physical qubits, so that quantum error correction is possible. A sketch of the QD potential and of the single-particle states is shown in Fig. \ref{eoverlaps}(a) for the vertically coupled QDs and in \ref{eoverlaps}(b) for the laterally coupled ones. In the insets, the relevant orbital electronic states are shown schematically: $|R\rangle$ and $|L\rangle$ are the states localized within each QD, and $|E\rangle$ is an extended state mediating the exchange interaction. The requirement that the qubits are uncoupled in their ground state (and addressable separately) while there exist transitions that allow for their coupling has been shown experimentally for vertical QDs;\cite{stinaff} in that case coupling is achieved in the excited state by an external electrical bias that brings a certain excited state in resonance, so that the electron of the optically generated exciton is in a molecular state of the two dots. For lateral dots, this requirement can be met if the potential along the line coupling the two QDs has the shape of an asymmetric mexican hat, an attractive potential with two asymmetric minima. The asymmetry could represent differences of size and composition between the two QDs, and it prevents the electrons in the ground states from tunneling into the other dot. As the energy quantum number is increased the states become more extended while still bound in the two QDs.
\\

The key component of our proposal is use of the electron-hole exchange interaction (EHEI), which has not been considered previously in conditional gate designs. EHEI gives rise to well-known features of QDs, such as the splitting between optically active (bright) and optically inactive (dark) excitons, as well as the splitting of the bright exciton doublet, which results in the linearly polarized photoluminescence of neutral QDs. These effects are crucial in the role of QDs as entangled photon emitters,\cite{akopian,greilich_entangled} and they have been studied extensively in recent years.\cite{urbaszek,finley,seguin,ediger,poem}

EHEI can play a significant role in optically controlled spin qubits, depending on the number of electrons in the intermediate optically excited state. For an even number of electrons the total electron spin can be zero, in which case there is no net EHEI. On the other hand, a state with an odd number of electrons always has half integer, i.e., nonzero total spin, which means nonzero net EHEI. Here we show that in the case of a two-spin conditional gate, where the intermediate state contains 3 electrons and one hole, the nonzero EHEI allows for a simple and fast design of a \textsc{c-phase} gate.

\section{Three-electron states}

The system we consider involves two resident electrons in two coupled QDs. The growth axis of the QDs defines the $z$ axis and is also the optical axis, and an external magnetic field along the perpendicular direction defines the $x$ axis (Voigt geometry). Upon optical excitation, the system consists of three electrons and one hole. The exchange interaction between the same quasi-particles (electron-electron) is the strongest interaction, on the order of 2-6 meV.\cite{scheibner} First we diagonalize the e-e exchange interaction in the three single particle basis states illustrated in Fig. \ref{eoverlaps}, $|L\rangle, |R\rangle, |E\rangle $. Then the states diagonalizing the e-e exchange Hamiltonian are eigenstates of the total electron spin $S$, which can be $3/2,1/2,1/2$, and the projection of that spin along the quantization axis.

The $S=3/2$ quadruplet is separable in spin and orbit, so it is straightforward to write it down as a product of the antisymmetric orbital state and the symmetric spin state: $|\mathcal{A}\rangle |{3}/{2},M_S\rangle$, where $M_S=\pm3/2,\pm1/2$ and $|\mathcal{A}\rangle=\frac{|LRE\rangle - |RLE\rangle - |LER\rangle + |REL\rangle - |ERL\rangle + |ELR\rangle }{\sqrt{6}}$. The energy of the $S=3/2$ states at zero $B$ field is $\epsilon_{3/2}=\epsilon_d-(V_{RE}+V_{LE})$, where $V_{RE}$ ($V_{LE}$) is the exchange integral between electrons in states $|E\rangle$ and $|R\rangle$ ($|L\rangle$), and $\epsilon_d$ is the energy of the three particles including direct Coulomb terms. The e-e exchange term between states $|R\rangle$  and $|L\rangle$ is essentially zero, and it has been dropped.

The remaining sets of states both have $S=1/2$. To find their energies and express them in terms of single particle states, we use the basis
states $|1_\uparrow\rangle = \frac{1}{\sqrt{6}}\left(|RLE\rangle|\uparrow\downarrow\uparrow\rangle - |LRE\rangle |\downarrow\uparrow\uparrow\rangle + c.p.
\right),~|2_\uparrow\rangle = \frac{1}{\sqrt{6}}\left(|RLE\rangle|\downarrow\uparrow\uparrow\rangle-|LRE\rangle |\uparrow\downarrow\uparrow\rangle
+ c.p.\right),~|3_\uparrow\rangle = \frac{1}{\sqrt{6}}\left(|RLE\rangle|\uparrow\uparrow\downarrow\rangle-|LRE\rangle |\uparrow\uparrow\downarrow\rangle + c.p.\right)$, (where $c.p.$ stands for cyclic permutations) for the $\sigma=+1/2$ states. The Coulomb interaction does not mix different spin projections, so we diagonalize each spin projection subspace separately. The Hamiltonian in the $|1_\uparrow\rangle, |2_\uparrow\rangle, |3_\uparrow\rangle$ basis is
\begin{eqnarray}
H_\uparrow = \left[\begin{array}{ccc}  \epsilon_d-V_{ER} & 0  &  -V_{EL} \\
0 & \epsilon_d-V_{EL} & -V_{ER}
\\
-V_{EL} & -V_{ER} & \epsilon_d
\end{array}\right].
\end{eqnarray}
Diagonalizing the above Hamiltonian gives the following eigenstates
\begin{eqnarray*}
|\frac{1}{2},\sigma;\pm\rangle = \frac{V_{LE}- V_{RE}\pm\epsilon}{-V_{LE}\mp\epsilon}|1_\uparrow\rangle+ \frac{V_{RE}}{-V_{LE}\mp\epsilon}|2_\uparrow\rangle+|3_\uparrow\rangle,
\end{eqnarray*}
with normalization $\sqrt{{  V_{LE}+ V_{RE} \pm 2\epsilon }/{\pm 6\epsilon}}$ (for the $\sigma=-1/2$ basis states all the spins are flipped). We label the states by their energies, $\pm\epsilon = \epsilon_d\pm\sqrt{V^2_{RE}+V^2_{LE}-V_{RE}V_{LE}}$, as $|1/2,\sigma;\pm\rangle$, where $\sigma=\pm 1/2$ is the spin projection along the quantization axis. In the following we take $ V_{LE}= V_{RE}\equiv J_{ee}$.

\section{Inclusion of electron-hole exchange interaction}

Now we consider the hole. The two heavy hole states in the QD act as a pseudospin and split in the presence of the magnetic field.\cite{xiaodong} Taking the tensor product of these with the three-electron states above gives a total of sixteen non-degenerate basis states. The exchange interaction between an electron labeled by $i$ and the hole is anisotropic, and can be written in terms of the hole pseudospin $j$ and the electron spin $s_i$ as \cite{bayer_exchange}
\begin{eqnarray}
H^{(i)}_{exch}= \sum_{\lambda=x,y,z} \alpha_\lambda(r_i) s_{i\lambda} j_\lambda ,
\label{ehexch}
\end{eqnarray}
where $\alpha$ is an operator acting on the envelope functions of the electron and the hole and is different for each of the principal axes.\cite{bayer_exchange} It can be shown that $\langle AB|\alpha_x|AB\rangle =-\langle AB|\alpha_y|AB\rangle =\Delta^{AB}_1 $, where $A,B$ are electron and hole envelope states respectively. These terms are responsible for the familiar anisotropic exchange between bright excitons in the $s$-shell. $\langle AB|\alpha_z|AB\rangle = \Delta^{AB}_0$ is responsible for the splitting between bright and dark excitons. The full exchange Hamiltonian is given by the sum of three terms of the form (\ref{ehexch}), one term for each electron. In the presence of a magnetic field, the full four-particle Hamiltonian can be expressed as
\begin{eqnarray}
H_{spin}
&=&\omega_e S_x + \omega_h j_x
\nonumber \\
&+&
\sum_{\lambda=x,y,z} \frac{\alpha_\lambda(r_1)+\alpha_\lambda(r_2)+\alpha_\lambda(r_3)}{3} S_\lambda j_\lambda
\nonumber\\
&+&
\frac{\alpha_\lambda(r_1)-\alpha_\lambda(r_2)}{3} (s_{1\lambda}-s_{2\lambda})j_\lambda + c.p..
\label{exchangeHam}
\end{eqnarray}
The terms in the first two lines conserve the total electron spin $S$. The terms in the last line are spin-antisymmetric and mix states of different total electron spin.

\begin{figure}[htp]
\begin{center}
\includegraphics[width=7cm,angle=-90]{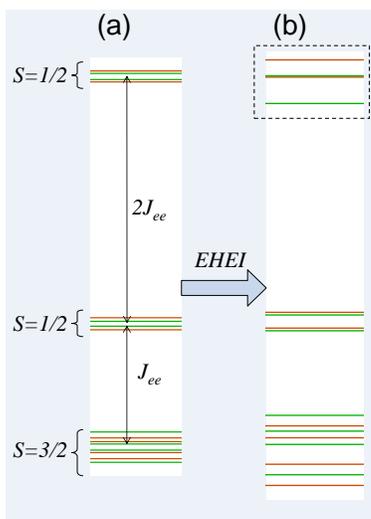}
\end{center}
\caption{(Color online) Three electron-one hole energy levels in the presence of the Voigt field (a) without and (b) with electron-hole exchange interaction. The total electron spin is approximately a good quantum number. The total parity is conserved (denoted here by color).}
\label{lvls}
\end{figure}

The Hamiltonian (\ref{exchangeHam}) is invariant under parity transformation, which means that $[H_{spin},R_x(\pi)]=0$, where $R_x(\pi)= e^{i\pi \mathcal{J}_x }$ is the operator that rotates the total ($\mathcal{J}=S+j$) spin about the quantization axis by angle $\pi$. The Hamiltonian is therefore block diagonal according to the parity eigenvalue: The states of even parity, \{$|\frac{3}{2},\frac{3}{2};h_x\rangle, |\frac{3}{2},\frac{\bar{1}}{2};h_x\rangle,|\frac{3}{2},\frac{1}{2};h_{\bar{x}}\rangle,  |\frac{3}{2},\frac{\bar{3}}{2};h_{\bar{x}}\rangle,|\frac{1}{2},\frac{1}{2};\pm;h_{\bar{x}}\rangle,\\|\frac{1}{2},\frac{\bar{1}}{2};\pm;h_x\rangle$\}, are labeled by $|\mathcal{E}\rangle$, and the states of odd parity, \{$|\frac{3}{2},\frac{3}{2};h_{\bar{x}}\rangle,  |\frac{3}{2},\frac{\bar{1}}{2};h_{\bar{x}}\rangle,|\frac{3}{2},\frac{1}{2};h_x\rangle, |\frac{3}{2},\frac{\bar{3}}{2};h_x\rangle,|\frac{1}{2},\frac{1}{2};\pm;h_x\rangle,\\|\frac{1}{2},\frac{\bar{1}}{2};\pm;h_{\bar{x}}\rangle$\}, are labeled by $|\mathcal{O}\rangle$. Therefore the Hamiltonian splits in two $8\times 8$ blocks. The energy levels are shown in panel (a) of Fig. \ref{lvls} without EHEI, while panel (b) includes EHEI, with typical values taken from experiment and Zeeman splittings corresponding to about 8 T. The orange color represents states of odd parity, while the green those of even parity. From the figure, one can see that the levels become mixed and shifted by the EHEI, with the mixing occurring predominantly within the same total electron spin multiplets. This is because the electron-electron exchange is much larger than the EHEI. It is therefore an appropriate approximation to ignore the spin-mixing terms in Eq. (\ref{exchangeHam}) and take $S$ to be a good quantum number. In the following we will focus on the topmost four states (dashed box in Fig. \ref{lvls}(b)), and use them for our gate design.

\begin{figure}[htp]
\begin{center}
\includegraphics[width=6.5cm,angle=-90]{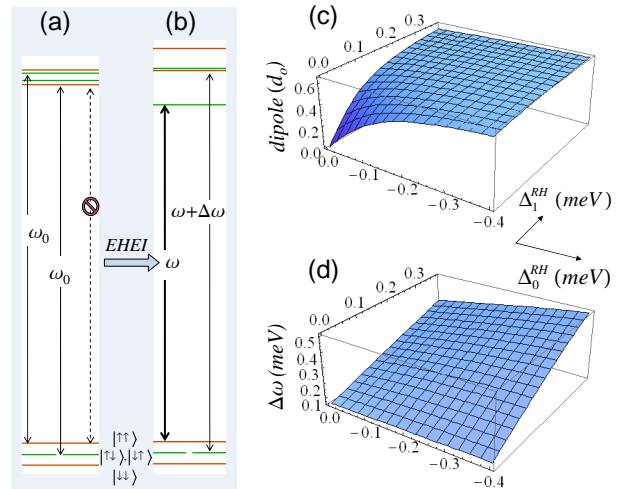}
\end{center}
\caption{(Color online) (a) In the absence of EHEI the optical transitions are degenerate and some transitions are forbidden by optical selection rules. (b) EHEI mixes the states and lowers the symmetry, such that all transitions are allowed. In particular, the lowest frequency transition indicated by the thick arrow can be used for a \textsc{c-phase} gate. (c) The dipole and (d) detuning of this transition as function of isotropic and anisotropic EHEI between state $|R\rangle$ and the hole state.}
\label{grouped}
\end{figure}

The Hamiltonian in this space is block diagonal, consisting of one block in the even subspace $\{|\frac{1}{2},\frac{1}{2};+;h_{\bar{x}}\rangle,~|\frac{1}{2},\frac{\bar{1}}{2};+;h_x\rangle \}$ and one in the odd subspace $\{|\frac{1}{2},\frac{1}{2};+;h_x\rangle,~|\frac{1}{2},\frac{\bar{1}}{2};+;h_{\bar{x}}\rangle \}$.  Diagonalizing each block gives
the four eigenenergies
\begin{eqnarray*}
E_{\mathcal{O}, \pm} &=& J_{ee} +  \Delta_1 \pm \sqrt{{(\omega_e+\omega_h)^2}/{4}+ ( \Delta_0+ \Delta_1)^2} \\
E_{\mathcal{E}, \pm} &=& J_{ee} -  \Delta_1 \pm \sqrt{(\omega_e-\omega_h)^2/4+ ( \Delta_0- \Delta_1)^2},
\end{eqnarray*}
where $\Delta_\kappa \equiv \frac{1}{6}(2\Delta^{RH}_\kappa + 2\Delta^{LH}_\kappa -\Delta^{EH}_\kappa)$, $\kappa=0,1$.
\\ \\
Now we also consider the $4\times4$ qubit subspace, shown at the bottom of Fig. \ref{grouped}(a) and (b), and corresponding to the two spins in the two QDs before optical excitation. The optical selection rules between this qubit subspace and the four states from Fig. \ref{lvls} are such that $H$-polarized light couples even to even and odd to odd states, whereas $V$ polarization couples odd with even states. These can be derived from the well-known circular selection rules if our basis states are expressed as linear combinations of the spin states polarized along the growth axis. They have also been demonstrated experimentally.\cite{xiaodong} First, let us ignore EHEI (Fig. \ref{grouped}(a)): then, not all ground states couple to every excited state. For example, the transition from the two-qubit state $|\uparrow\uparrow\rangle$ to the lowest state of the quadruplet, $|\frac{1}{2},\frac{\bar{1}}{2};+\rangle |h\rangle$ is optically forbidden, as indicated in Fig. \ref{grouped}(a). To implement a \textsc{c-phase} gate optically, a transition has to be addressed between $|\uparrow\uparrow\rangle$ (or $|\downarrow\downarrow\rangle$) and an excited state, without other transitions being excited. However, in the absence of EHEI, each transition has a frequency and polarization that is doubly degenerate. An example is shown in Fig. \ref{grouped}(a), for the two transitions labeled by their common frequency, $\omega_0$. This means that the target transition cannot be isolated, irrespective of the laser bandwidth and polarization.
\\ \\
Now, we include EHEI, in panel (b) of Fig. \ref{grouped}. The transition between $|\uparrow\uparrow\rangle$ and the lowest excited state now is allowed, and moreover it has a unique frequency. This transition is \emph{enabled by} the EHEI. Stronger EHEI results in a larger dipole moment for this transition and larger energy separation from the nearest transition. This can be seen in Fig. \ref{grouped}, panels (c) and (d), where the dipole and detuning $\Delta\omega$ are plotted as functions of the isotropic and anisotropic EHEI between the hole state and the electronic state $|R\rangle$, $\Delta^{RH}_0$ and $\Delta^{RH}_1$. Here the hole has been taken to be confined within one QD (the one with state $|R\rangle$), so the EHEI with state $|L\rangle$ is zero; for state $|E\rangle$ we have taken $\Delta^{EH}_0=\Delta^{RH}_0/2$ and $\Delta^{EH}_1 = 0$. The range of values for the EHEI shown in Fig. \ref{grouped}(c),(d) is taken from experiments.\cite{bayer_exchange,finley,scheibner,poem}

\section{Two-qubit c-phase gate}

For the 2-qubit gate, we propose using a laser of pulse area 2$\pi$ resonant with the transition discussed above (thick arrow in Fig. \ref{grouped}(b)). The pulse can have any temporal shape, and its effect will be to induce a minus sign to the $|\uparrow\uparrow\rangle$ state, leaving the rest of the basis states unaffected, i.e. it can be expressed by the unitary $diag\{-1,1,1,1\}$. This is the $\textsc{c-phase}$ quantum gate, which we denote by $C$. The result that inclusion of EHEI makes possible this 2-qubit gate is remarkable, since EHEI is usually viewed as a source of decoherence and error.

Our $\textsc{c-phase}$ gate is compatible with the single qubit gates that we developed earlier,\cite{economou_rotations} since the two QDs have optical excitations at different frequencies (about 10 meV apart), so the speed of the single qubit gates is not compromised. These single spin gates, implemented through the lowest trion localized in each QD, in combination with the present $\textsc{c-phase}$ gate, provide the entangling \textsc{cnot} gate as $R^\dag _z(\frac\pi 2)C R_z(\frac\pi 2)$, where $R_n(\phi)$ is the single qubit operator representing a rotation by angle $\phi$ about axis $\hat{n}$. Single qubit rotations about $z$ have picosecond durations,\cite{economouprb06} so the total gate time is not slowed down by the single-qubit gates.

\begin{figure}[htp]
\begin{center}
\includegraphics[width=3.4cm,angle=270]{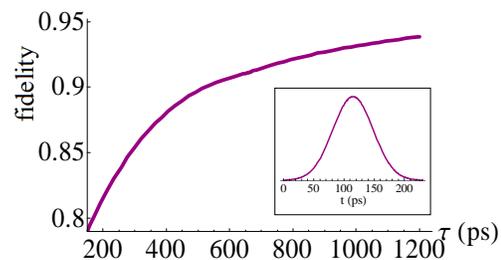}
\end{center}
\caption{(Color online) Fidelity as function of the effective decay time of the optically excited state for a gaussian pulse envelope, shown in the inset.}
\label{fid}
\end{figure}

We have performed numerical calculations in the $8\times8$ space of Fig. \ref{grouped}(b) to simulate the \textsc{c-phase} gate in the presence of decoherence and relaxation dynamics. The fidelity has been calculated as the average \cite{emerson}
\begin{eqnarray}
f = \frac{1}{10}\sum_i |I_{ii}|^2 + \frac{1}{20}\sum_{i\neq k} \left( I_{ii}I^*_{kk} + I_{ik}I^*_{ik} \right),
\end{eqnarray}
where $I=U^\dag U_o$, $U_o$ is the ideal operation, and $U$ is the actual operation. The indices $i,k$ run through all four possible two-qubit states. The predominant dissipation mechanisms are relaxation to lower levels and spontaneous emission of the trion (recombination). We have used an effective recombination rate $\tau$ in the Lindblad master equations to account for both of these mechanisms. In those simulations we used the following values for the parameters: for the Zeeman splittings $\omega_e =0.12$ meV, $\omega_h =0.04 $ meV, which corresponds to a $B$ field of about 8 T for InAs QDs. For the EHEI we used the recently measured values by Poem \textsl{et al.},\cite{poem} $\Delta^{RH}_0=-0.2$ meV, $\Delta^{RH}_1= 0.38$ meV, $\Delta^{EH}_0=-0.1 $ meV, $\Delta^{EH}_1= \Delta^{LH}_0= \Delta^{LH}_1= 0$. We have used a gaussian pulse envelope, $e^{-\sigma^2 t^2/\hbar^2 }$, with $\sigma=0.01$ meV, shown in the inset of Fig. \ref{fid}. Fig. \ref{fid} shows the calculated fidelity for varying values of $\tau$: the fidelity ranges from 79\% for $\tau\simeq 160$ ps to about 94\% for $\tau \simeq 1200$ ps.
We note that the values of the EHEI used in our calculations were chosen for concreteness, and our proposal is valid for a wide range of EHEI values. Stronger EHEI is advantageous, as it better isolates the target transition from the other transition, Fig. \ref{grouped}(d), leading to higher fidelity and shorter gate times.
\\ \\
\section{Conclusions}

In conclusion, we have proposed an optically induced $\textsc{c-phase}$ gate between spins in neighboring (coupled) QDs. This gate, in combination with our single qubit rotations, is equivalent to the familiar \textsc{c-not} gate. Our proposal yields fidelities on the order of 85\% for gates of pulsewidth of about 100 ps.  The electron-hole exchange interaction has been shown to be a crucial component of this conditional control by isolating a transition in the system, thereby making the gate possible.

\begin{acknowledgments}
We thank S.C. Badescu for useful discussions. This work is supported by the US Office of Naval Research. One of
us (S.E.E.) is an NRC/NRL Research Associate.
\end{acknowledgments}

\end{document}